\documentclass[aps,prb,twocolumn,showpacs,superscriptaddress,10pt,longbibliography]{revtex4-2}

\usepackage{amssymb}
\usepackage{graphicx}
\usepackage[colorlinks = true,linkcolor = red,citecolor = blue,urlcolor=blue]{hyperref}
\usepackage[sort&compress]{natbib}
\usepackage{scalerel}
\usepackage[normalem]{ulem}
\usepackage[dvipsnames]{xcolor}
\usepackage{bm}
\usepackage{amsmath}
\usepackage{subfigure}
\usepackage{amsfonts}
\usepackage{mathtools}
\usepackage{dsfont}
\usepackage{float}
\usepackage{lmodern}
\usepackage[many]{tcolorbox}
\usepackage{empheq}
\usepackage{soul}
\usepackage[T1]{fontenc}

\newcommand{\psitgt}{\ensuremath{\psi_{\text{target}}}}

\DeclarePairedDelimiter\ket{|}{\rangle}
\DeclarePairedDelimiterX\braket[2]{\langle}{\rangle}{#1 \delimsize\vert #2}

\begin{document}
\title{Preparation of a Laughlin-like fractional Chern insulator on a superconducting quantum processor}

\author{Kamil Dutkiewicz}
\affiliation{DIPC - Donostia International Physics Center, Paseo Manuel de Lardiz{\'a}bal 4, 20018 San Sebasti{\'a}n, Spain}
\affiliation{Department of Electricity and Electronics, FCT-ZTF, University of the Basque Country (EHU), Leioa, Spain}

\author{B\'{a}rbara Andrade}
\affiliation{LaBRI - Laboratoire Bordelais de Recherche en Informatique, Universit\'e de Bordeaux, CNRS, 33405 Talence, France}
\affiliation{IFT-UNESP Instituto de F\'isica Te\'orica, Universidade Estadual Paulista, ICTP-SAIFR South American Institute for Fundamental Research, Rua Dr.~Bento Teobaldo Ferraz 271, Bloco II, Barra Funda, 01140-070 S\~ao Paulo, Brazil}

\author{Geza Giedke}
\affiliation{DIPC - Donostia International Physics Center, Paseo Manuel de Lardiz{\'a}bal 4, 20018 San Sebasti{\'a}n, Spain}
\affiliation{IKERBASQUE, Basque Foundation for Science, Plaza Euskadi 5, 48009 Bilbao, Spain}
\author{Tobias Grass}
\affiliation{DIPC - Donostia International Physics Center, Paseo Manuel de Lardiz{\'a}bal 4, 20018 San Sebasti{\'a}n, Spain}
\affiliation{IKERBASQUE, Basque Foundation for Science, Plaza Euskadi 5, 48009 Bilbao, Spain}

\begin{abstract}
Fractional Chern insulators (FCIs) are lattice analogues of fractional quantum Hall states, hosting topological order and fractionalized anyonic excitations. Their realization poses a demanding target for quantum simulation. Here, we realize a bosonic FCI on an IBM Nighthawk superconducting quantum processor using classically optimized variational circuits. We target a Laughlin-like state that is an exact zero-energy ground state of a long-range parent Hamiltonian on a torus, eliminating physical edges and providing a favorable geometry for accessing fractional quantum Hall physics in a small system. On a $4\times4$ qubit lattice, we prepare both the four-boson FCI ground state and a three-boson state hosting two quasiholes. Measurements of particle densities and correlations reproduce the characteristic structure of the target states, while reconstructed chiral magnetization currents resolve the circulating current pattern associated with the quasiholes. The preparation of these strongly correlated topological states with shallow circuits establishes a route toward the digital simulation and manipulation of fractionalized excitations, including measurements of topological invariants and anyonic braiding.
\end{abstract}

\maketitle

\section{Introduction}
Since the discovery of the fractional quantum Hall (FQH) effect~\cite{Tsui82,Laughlin83}, topological quantum matter has been prized for hosting exotic anyonic quasiparticles with fractional statistics~\cite{Leinaas77,Wilczek82}, offering intrinsically protected carriers with potential applications for fault-tolerant quantum computing~\cite{Nayak08}. While FQH states arise naturally in two-dimensional electron gases under strong magnetic fields, synthetic realizations are essential for practical harnessing of their properties. Analog platforms -- including rotating atomic traps~\cite{Cooper01,Regnault03,Cooper08}, ultracold atoms with synthetic gauge fields~\cite{Julia-Diaz11, Aidelsburger18}, and optical cavities -- have demonstrated atomic Laughlin states~\cite{Gemelke10,Lunt24} and polaritonic FQH phases~\cite{Clark20}. However, reaching the FQH regime remains challenging due to the need for large angular momentum and fragile adiabatic state preparation across small many-body gaps~\cite{Popp04,Roncaglia11,Andrade21}.
Digital quantum computers offer a complementary route~\cite{Latorre10,Kaicher20,Rahmani20,Wu26}, though experimental progress has been limited to fermionic Laughlin states on seven- to sixteen-qubit processors~\cite{Kirmani22,Kirmani23,Shen26,Exposito26}.

Fractional Chern insulators (FCIs) are lattice counterparts of FQH states~\cite{Sorensen05, Hafezi07,Regnault11}, realizing fractionalized topological phases in partially filled Chern bands without requiring a continuum Landau level. They exhibit the characteristic phenomenology of FQH matter, including a gapped incompressible bulk, topological ground-state structure, and fractionalized excitations~\cite{Repellin20,Wang22, Lu25}. In contrast to continuum Laughlin states, however, generic FCI wave functions do not admit simple analytical forms. Important exceptions are provided by specially engineered parent Hamiltonians~\cite{Kapit10,Nielsen13}, whose ground states can be related exactly to continuum FQH states.
Although synthetic gauge fields and complex hopping amplitudes have enabled the realization of topological single-particle bands in optical lattices~\cite{Aidelsburger13,Miyake13}, extending these ideas to strongly correlated FCIs is substantially more demanding, and a range of preparation strategies has therefore been proposed~\cite{Grusdt14,Blatz24,Wu25, Liu26}.
Recent experiments have realized FCI-like states in small optical-lattice systems~\cite{Leonard23,Kwan26} and superconducting qubit arrays~\cite{Wang24}. Related topologically ordered states, including chiral spin liquids and $\mathbb{Z}_2$ spin liquids, have also been prepared on trapped-ion, Rydberg-atom, and superconducting-qubit platforms 
\cite{Iqbal24,Semeghini21,Satzinger21,Xu23}. These advances highlight the growing ability of programmable quantum systems to access strongly correlated topological matter, while the digital preparation of a Laughlin-like FCI ground state remains an important challenge.

In this work, we report the digital preparation of a bosonic FCI in the Laughlin phase on an IBM Nighthawk superconducting quantum processor. We target the ground state of the Kapit--Mueller Hamiltonian on a $4\times4$ square lattice, whose fine-tuned long-range complex hopping realizes an exactly flat Chern band closely reproducing the structure of a continuum Landau level~\cite{Kapit10}. 
Periodic boundary conditions remove physical edges and provide a particularly favorable setting for accessing Laughlin physics in a small system. At flux density $1/2$ per plaquette, we prepare the four-boson Laughlin-like ground state using a classically optimized variational circuit with four steps and $264$ CZ gates (in $44$ 2-qubit gate layers), reaching a fidelity of $0.95$ in noiseless simulations. Mitigating hardware errors by particle-number post-selection and dynamical decoupling, the experimental realization yields densities, Fock-state populations, and correlation observables in good quantitative agreement with the target state. We further prepare a three-boson state containing two quasiholes and resolve the associated chiral magnetization currents. These results establish a digital route to preparing and probing fractional Chern insulator states and provide a basis for future measurements of topological invariants and anyonic braiding~\cite{Kapit12}.

\begin{figure*}[t]
\centering    \includegraphics[width=0.9\textwidth]{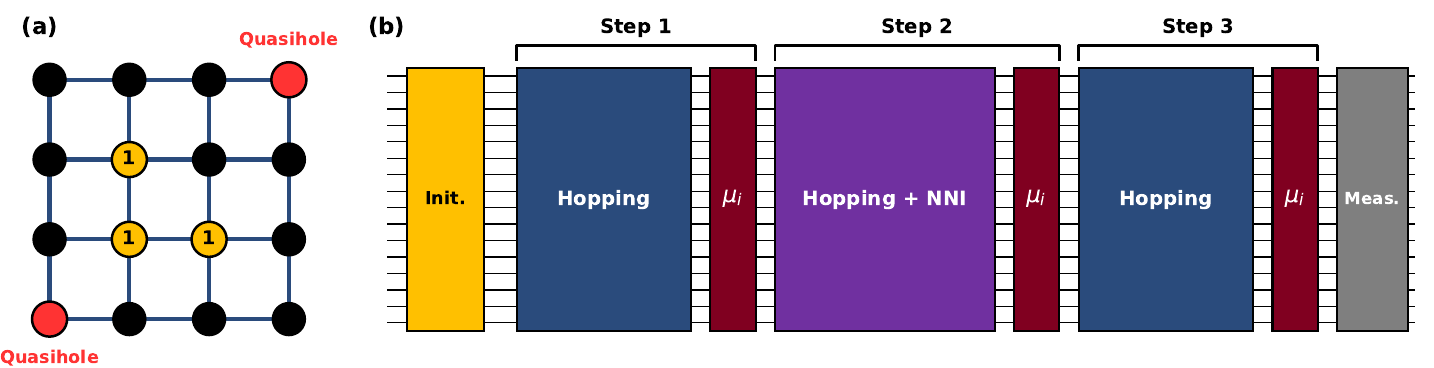}
    \caption{\textbf{State-preparation protocol for the $N=3$ quasihole state.}
(a) Initial product state relative to the quasihole positions. Qubits on yellow sites are initialized in $\ket{1}$, while all others are initialized in $\ket{0}$. Red sites indicate the quasihole pinning positions in the target state.
(b) Variational state-preparation circuit for $N=3$ particles and two quasiholes. The optimized protocol reaches a noiseless fidelity $F=0.95$ using $p=3$ steps. In each step $q$, nearest-neighbor hopping with complex amplitude $J^q_{ij}$ and on-site potentials $\mu^q_i$ are applied for unit time, with independent parameters for each step optimized jointly to maximize the overall fidelity $F$. Selected steps additionally include nearest-neighbor interactions $V^q_{ij}$, increasing the implementation cost from two to three CZ gates per bond.}
    \label{fig:circuitProtocol}
\end{figure*}

\section{Results}
\subsection{State preparation}

We map the qubits onto the sites of a lattice of hard-core bosons and target the bosonic Laughlin state at filling factor \(\nu=1/2\). Its lattice version is provided by the Kapit--Mueller
Hamiltonian~\cite{Kapit10} (see Methods Sec.~\ref{app:km_ham}). Through fine-tuned long-range complex hopping amplitudes, this model reproduces the flat-band structure of the lowest Landau level on a lattice. For a magnetic flux \(\phi=1/2\) per plaquette, an \(L\times L\) lattice with periodic boundary conditions contains \(N_\phi=L^2/2\) flux quanta. At filling \(\nu=N/N_\phi=1/2\), corresponding to \(N=\frac{L^2}{4}\) hard-core bosons, the Laughlin-like FCI is an exact zero-energy ground state. On the torus, the ground-state manifold is twofold degenerate.

Localized quasiholes can be introduced by adding repulsive local potentials and reducing the particle number. Removing one boson creates two quasiholes in the \(\nu=1/2\) Laughlin phase, such that a state with \(N_{\rm qh}\) quasiholes contains
\(
N=\frac{L^2}{4}-\frac{N_{\rm qh}}{2}
\)
bosons. We focus on a \(4\times4\) lattice and consider either the \(N=4\) Laughlin-like ground-state manifold or an \(N=3\) state containing two quasiholes.

For both cases, we numerically determined the state vectors spanning the degenerate target manifold by exact diagonalization. Denoting these orthonormal states by \(\ket{\psi_{{\rm target},\alpha}}\), with \(\alpha=1,2\), we then classically optimized a shallow variational circuit to maximize its fidelity with the target manifold, quantified by
\begin{equation}
F=\sum_{\alpha=1}^{2}
\left|\braket{\psi_{{\rm target},\alpha}}{\psi_{\rm sim}}\right|^2,
\label{eq:fidelity}
\end{equation}
where \(\ket{\psi_{\rm sim}}\) is the noiseless circuit output.

The preparation protocol is sketched in Fig.~\ref{fig:circuitProtocol}. For the \(N=3\) quasihole state, panel (a) shows the initial product state, with \(N\) qubits initialized in \(\ket{1}\) and the remaining qubits in \(\ket{0}\). The two pinning potentials defining the quasihole positions are placed at opposite corners of the lattice. As illustrated in panel (b), the circuit consists of \(p=3\) variational steps. In each step \(q\), we implement a first-order Trotterized evolution generated by
\begin{equation}
H^q =
\sum_{\langle i,j\rangle}
\left[
J^q_{ij} a_i^\dagger a_j + {\rm h.c.}
+ V^q_{ij} n_i n_j
\right]
+
\sum_i \mu^q_i n_i .
\label{eq:StepHam}
\end{equation}
Here \(J^q_{ij}\) denotes a complex nearest-neighbor hopping amplitude, \(V^q_{ij}\) a nearest-neighbor interaction, and \(\mu^q_i\) an on-site potential. Each variational step has its own set of parameters \(J^q_{ij}\), \(V^q_{ij}\), and \(\mu^q_i\), and the parameters of all steps are optimized jointly to maximize the manifold fidelity $F$; see also Methods Sec.~\ref{app:jax}. A hopping operation between two neighboring sites requires two CZ gates, whereas a combined hopping and interaction operation requires three CZ gates [Fig.~\ref{fig:circuitAppendix}(d,e)]. To limit the circuit depth for $N=3$, nonzero \(V^q_{ij}\) are included only in the second variational step. The \(N=4\) ground state is prepared using an analogous protocol with \(p=4\) steps [Fig.~\ref{fig:circuitAppendix}(a,b)]. The first-order decomposition of Eq.~\eqref{eq:StepHam} is shown in Fig.~\ref{fig:circuitAppendix}(c). Importantly, this Trotterized gate structure is incorporated directly into the classical optimization, so that Trotter errors are absorbed into the optimized variational parameters rather than introduced after optimization.

The circuit architecture combines physically motivated local operations with variational optimization, while preserving total particle number exactly. The variational parameters are optimized to prepare the target state directly, without requiring an explicit dynamical path. This approach enables a shallow circuit composed of on-site and nearest-neighbor operations to approximate states in the ground-state manifold of the long-range Kapit--Mueller Hamiltonian, without explicitly implementing its long-range hopping terms.

The optimized \(p=3\) and \(p=4\) circuits both reach \(F=0.95\) in noiseless simulations. For \(N=3\), the circuit contains 168 CZ gates with a 2-qubit-gate depth of 28; for \(N=4\), it contains 264 CZ gates with a 2-qubit-gate depth of 44. We executed both protocols on a \(4\times4\) patch of the 120-qubit \texttt{ibm\_miami} Nighthawk R1 processor. Since the ideal circuit conserves total particle number, we mitigated bit-flip errors by post-selecting measurement outcomes with the correct particle number. For \(N=3\), \(80{,}000\) measured shots yielded a retention rate of \(34.6\%\), while for \(N=4\), \(60{,}000\) shots yielded a retention rate of \(33.3\%\).

\subsection{State detection}
\subsubsection{Density profiles}

\begin{figure*}[t]
\centering
\includegraphics[width=0.99\textwidth]{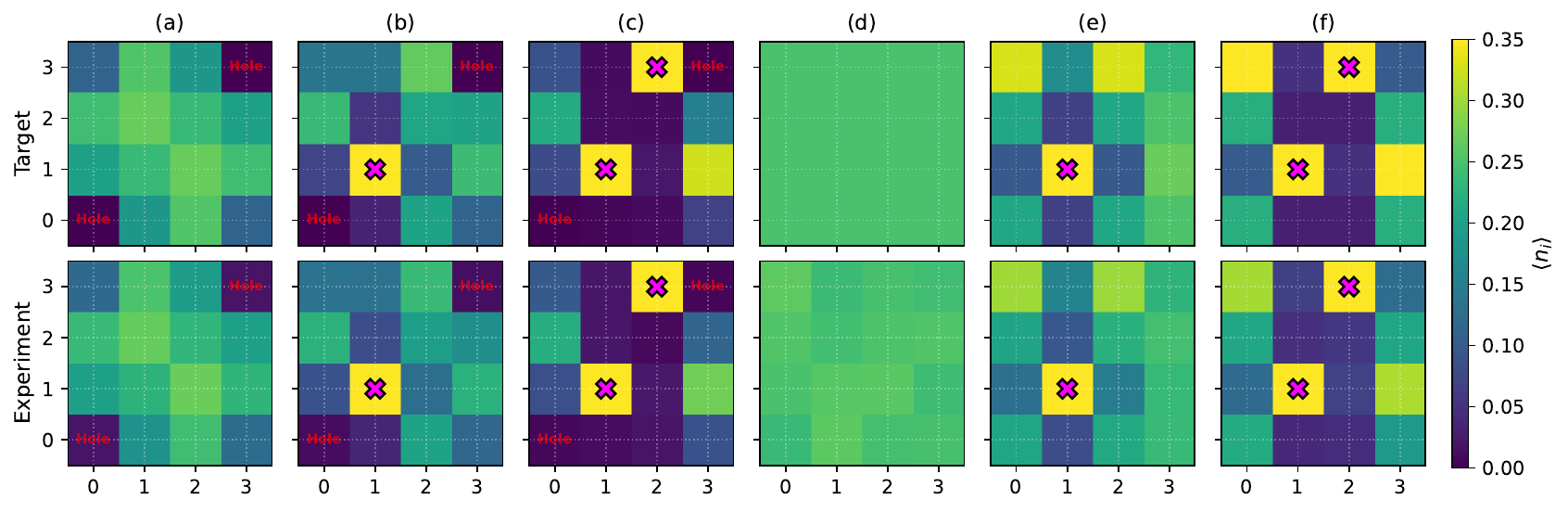}
  \caption{\textbf{Local densities and conditional spatial correlations.}
The top row shows results for the target state $\ket{\psi_{\rm target}}$, while the bottom row shows post-selected experimental data. Panels (a)--(c) correspond to $N=3$ with two quasiholes pinned at $(0,0)$ and $(3,3)$ (red markers), and panels (d)--(f) to the $N=4$ ground state without quasiholes. Local densities $\langle n_i\rangle$ are shown in (a,d). Panels (b,e) show conditional densities given a particle at $(1,1)$ (magenta marker), while (c,f) show conditional densities given particles at both $(1,1)$ and $(2,3)$. In (c), the density at $(0,3)$ is strongly suppressed by its proximity to the quasihole pinning sites across the magneto-periodic boundary.}
    \label{fig:Densities}
\end{figure*}

Using the post-selected measurement outcomes, we reconstructed the local particle densities \(\langle n_i\rangle\), shown in Figs.~\ref{fig:Densities}(a) and \ref{fig:Densities}(d). For \(N=3\), the density is suppressed at the two quasihole pinning sites, \((0,0)\) and \((3,3)\), whereas the \(N=4\) state shows an approximately uniform density of \(\langle n_i\rangle \simeq 0.25\) across the \(16\) sites. In both cases, the experimental data show strong quantitative agreement with the target state \(\ket{\psi_{\rm target}}\), defined as the projection of \(\ket{\psi_{\rm sim}}\) onto the target manifold,
\begin{equation}
\ket{\psi_{\rm target}} \equiv
\frac{
\sum_{\alpha}
\ket{\psi_{{\rm target},\alpha}}
\braket{\psi_{{\rm target},\alpha}}{\psi_{\rm sim}}
}{
\|
\sum_{\alpha}
\ket{\psi_{{\rm target},\alpha}}
\braket{\psi_{{\rm target},\alpha}}{\psi_{\rm sim}}
\|
}.
\label{eq:psitgt}
\end{equation}

To probe spatial correlations, we evaluated conditional densities for representative particle configurations. Conditioning on a particle at \((1,1)\) [Figs.~\ref{fig:Densities}(b,e)] reveals a pronounced suppression of density on neighboring sites. Conditioning on a second particle at \((2,3)\) [Figs.~\ref{fig:Densities}(c,f)] further confines the remaining density to sites that maximize particle separation. For \(N=3\) [Fig.~\ref{fig:Densities}(c)], the quasiholes additionally suppress the density at \((0,3)\), which is adjacent to the pinning sites across the magneto-periodic boundary. Overall, the measured conditional distributions closely match those calculated for the target states, apart from a small, approximately uniform noise background.

\subsubsection{Particle distributions}
To characterize the prepared states beyond local densities, we analyzed the post-selected probability distributions over particle configurations. For the \(4\times4\) lattice, the fixed-particle-number Hilbert spaces contain \(\binom{16}{3}=560\) configurations for \(N=3\) and \(\binom{16}{4}=1820\) for \(N=4\).

For \(N=4\), the configuration probabilities of the target state \psitgt\, are translationally invariant and possess twofold rotational (\(C_2\)) symmetry. The particular state selected within the two-dimensional ground-state manifold does not retain \(C_4\) symmetry. Grouping configurations related by translations and the remaining point-group symmetries reduces the \(1820\) configurations to \(78\) equivalence classes. For \(N=3\), the pinned quasiholes break translational symmetry, leaving \(280\) equivalence classes.

\begin{figure*}[t]
\centering
    \includegraphics[width=0.99\textwidth]{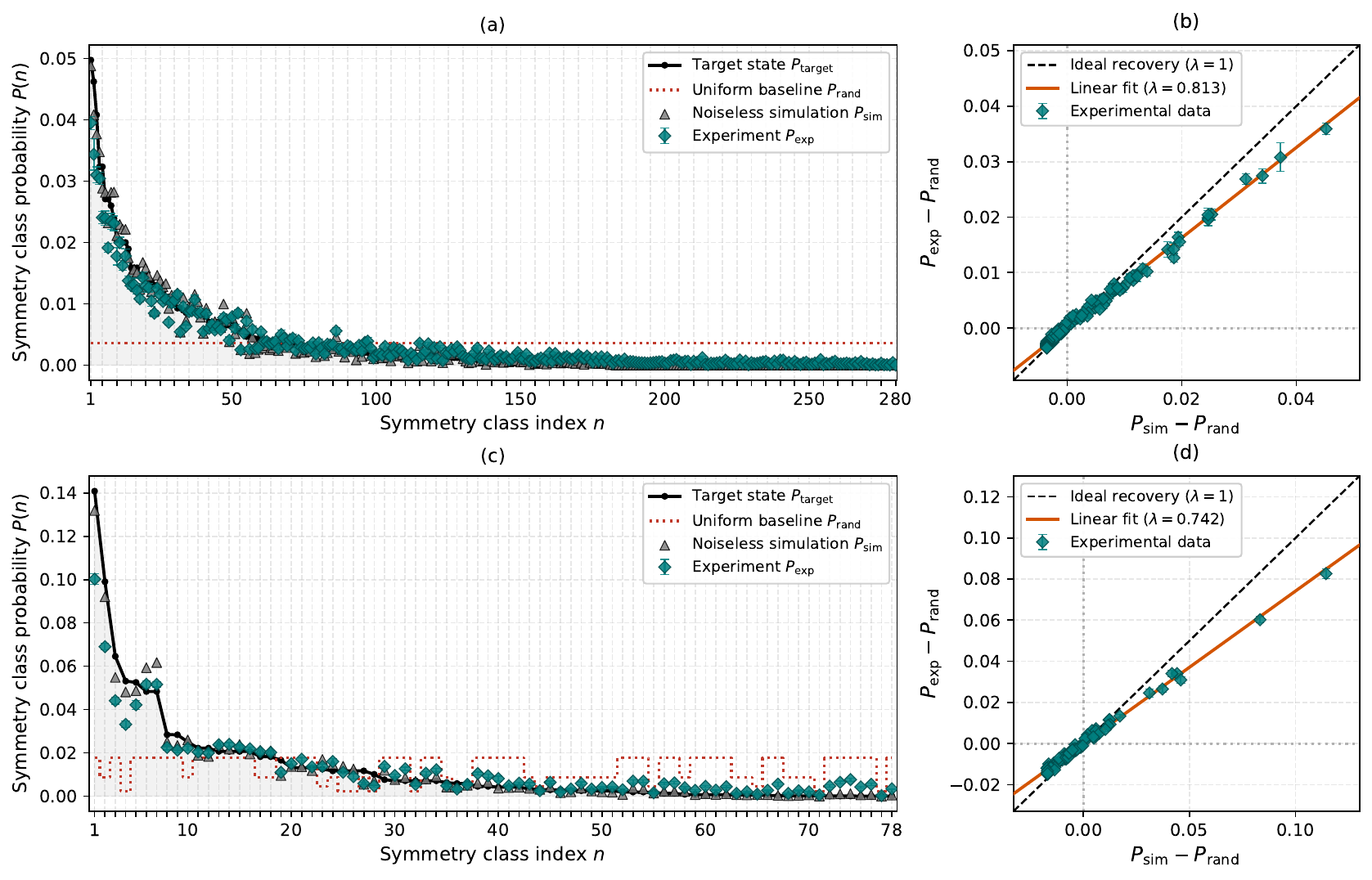}
    \caption{\textbf{Particle-configuration probability distributions.}
Panels (a,b) show the $N=3$ state with two quasiholes ($27{,}661$ post-selected outcomes from $80{,}000$ shots), and panels (c,d) the $N=4$ ground state ($19{,}965$ post-selected outcomes from $60{,}000$ shots). In (a,c), particle configurations are grouped into equivalence classes according to the spatial symmetries of the corresponding system and ordered by their target-state probability $P_{\rm target}$. Experimental probabilities $P_{\rm exp}$ (green diamonds) are compared with the target state $\ket{\psi_{\rm target}}$ (black line), the noiseless circuit simulation $P_{\rm sim}$ (grey triangles), and the maximally mixed fixed-particle-number baseline $P_{\rm rand}$ (red dotted line). Panels (b,d) show $P_{\rm exp}-P_{\rm rand}$ versus $P_{\rm sim}-P_{\rm rand}$. Dashed black lines indicate ideal agreement, $P_{\rm exp}=P_{\rm sim}$, while solid orange lines show linear fits with slopes $\lambda=0.813$ for $N=3$ and $\lambda=0.742$ for $N=4$ and negligible intercepts. Error bars denote the standard deviation across eight physical-to-logical qubit mappings.}
    \label{fig:populationDistribution}
\end{figure*}

Figures~\ref{fig:populationDistribution}(a,c) compare the class probabilities of the experimental data, \(P_{\rm exp}\), and noiseless circuit simulations, \(P_{\rm sim}\), with those of the target state. To average over spatial variations across the processor, each circuit was executed using eight physical-qubit mappings corresponding to rotations and reflections of the virtual \(4\times4\) lattice; error bars denote the standard deviation across these realizations. We also show the probability distribution \(P_{\rm rand}\) of a maximally mixed state within the corresponding fixed-\(N\) sector. Because the equivalence classes have different degeneracies, \(P_{\rm rand}\) is not uniform across classes. As expected from the high simulated preparation fidelity, \(P_{\rm sim}\) closely follows the target distribution, while the experimental probabilities retain the same structure but are systematically shifted toward \(P_{\rm rand}\).

To quantify this attenuation, Figs.~\ref{fig:populationDistribution}(b,d) show \(P_{\rm exp}-P_{\rm rand}\) versus \(P_{\rm sim}-P_{\rm rand}\). Linear fits yield slopes \(\lambda=0.813\) for \(N=3\) and \(\lambda=0.742\) for \(N=4\), with the data closely following the linear trend. For purely uncorrelated noise, \(\lambda\) could be interpreted directly as the fraction of error-free realizations; the noise in the experimental data, however, retains nontrivial spatial correlations. A separate noise-scaling analysis (see Methods Sec.~\ref{app:noise_scaling}) estimates error-free fractions of \(60(2)\%\) for \(N=3\) and \(46(5)\%\) for \(N=4\) after post-selection. The configuration distributions therefore show that the characteristic structure of the target states survives on hardware, with predominantly noise-induced attenuation toward the mixed-state background.

\subsubsection{Particle current measurements}

A characteristic consequence of the broken time-reversal symmetry of the FCI state is the appearance of chiral equilibrium currents around localized quasiholes. 
To probe the chiral character of the prepared quasihole state, we reconstructed the expectation values of the bond-current operators associated with the Kapit--Mueller Hamiltonian. Although the preparation circuit does not implement the Kapit--Mueller dynamics directly, the corresponding equilibrium current pattern can be obtained from the measured single-particle coherences of the prepared state. In the target FCI state, the quasihole-induced density depletion is accompanied by a circulating magnetization current. Because of the twofold ground-state degeneracy, the detailed bond currents depend on the particular superposition within the target manifold. However, the net circulation has the same sign throughout the entire manifold, with
$
-0.342 \le I_{\rm total} \le -0.234,
$
demonstrating a robust chirality of the quasihole current. 

\begin{figure*}[t]
\centering    \includegraphics[width=0.99\textwidth]{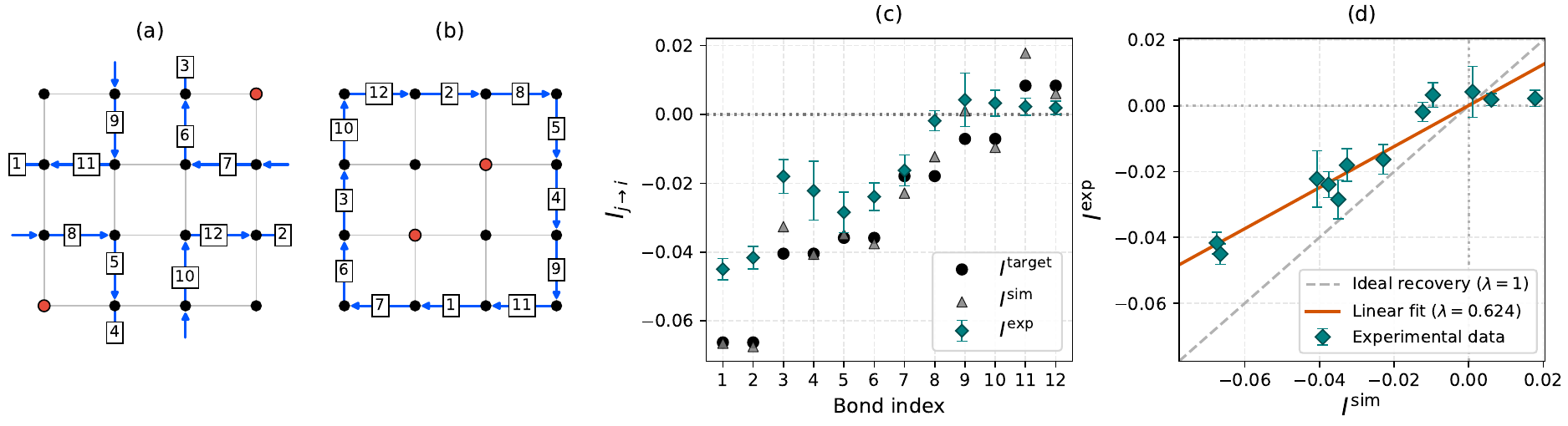}
    \caption{\textbf{Reconstructed bond currents around the quasiholes.}
(a) $4\times4$ lattice showing the directed bonds along which the Kapit--Mueller current expectation values are reconstructed. (b) Equivalent representation shifted by $(2,2)$ under periodic boundary conditions, illustrating that the directed bond path forms a closed loop around the quasiholes. (c) Experimental bond currents $I^{\rm exp}$ (green diamonds), compared with noiseless circuit results $I^{\rm sim}$ (grey triangles) and target-state values $I^{\rm target}$ (black circles) for the $12$ bonds, ordered by increasing $I^{\rm target}$. Each bond current was reconstructed from $20{,}000$ shots before post-selection. (d) Experimental currents $I^{\rm exp}$ versus noiseless circuit values $I^{\rm sim}$. The black dashed line denotes ideal agreement, $I^{\rm exp}=I^{\rm sim}$, and the solid orange line is a linear fit with slope $\lambda=0.624$ and intercept $b=10^{-4}$. Error bars denote the standard deviation across eight physical-qubit mappings.}
    \label{fig:Current}
\end{figure*}

We probed this circulation along the directed loop shown in Figs.~\ref{fig:Current}(a,b). The particle current along a bond from site $j$ to site $i$ is
\begin{equation}
    I_{j\to i}
    =
    2\,\operatorname{Im}
    \left[
    J_{ij}\langle a_i^\dagger a_j\rangle
    \right],
\end{equation}
where $J_{ij}$ is the corresponding Kapit--Mueller hopping amplitude. The coherence $\langle a_i^\dagger a_j\rangle$ was reconstructed experimentally from 2-qubit measurements in the $X$ and $Y$ bases, while particle-number post-selection was used to suppress errors. Details of the measurement protocol and the conversion from conditional to physical bond currents are given in the Methods section, see~Sec.~\ref{app:currents}.

Figure~\ref{fig:Current}(c) compares the measured currents $I^{\rm exp}$ with the noiseless circuit values $I^{\rm sim}$ and the projected target-state values $I^{\rm target}$ along all $12$ bonds of the loop. The experimental profile reproduces the spatial structure and direction of the simulated current, although with an overall attenuation toward zero. This suppression is consistent with decoherence and depolarizing noise, which reduce the off-diagonal coherence entering the current operator. As shown in Fig.~\ref{fig:Current}(d), $I^{\rm exp}$ remains approximately linear in $I^{\rm sim}$, with a fitted slope $\lambda=0.624$.

Summing the bond currents around the loop gives a net experimental circulation
$$
I_{\rm total}^{\rm exp}=-0.19\pm0.06,
$$
compared with
$$
I_{\rm total}^{\rm sim}=-0.300,
\qquad
I_{\rm total}^{\rm target}=-0.319.
$$

\section{Discussion}
In summary, we realized Laughlin-like fractional Chern insulator states on a digital quantum processor using shallow circuits obtained through classical variational optimization. The prepared states reproduced the local densities and spatial particle correlations of the target Kapit--Mueller ground-state manifold, with deviations predominantly appearing as an attenuation toward a mixed-state background. For the $N=3$ state containing two quasiholes, we additionally reconstructed the Kapit--Mueller bond-current expectation values and observed a finite chiral circulation around the quasiholes, with the direction and spatial structure consistent with theoretical predictions. These results demonstrate that gate-based quantum processors can prepare and probe strongly correlated lattice states with fractional quantum Hall character using circuits of experimentally accessible depth.

A central ingredient is the particle-number-conserving variational ansatz, which combines a flexible parametrization with substantial gate parallelism. Besides reducing circuit depth, exact $U(1)$ symmetry enables efficient mitigation of particle-loss and bit-flip errors through global particle-number post-selection. The optimized $p=3$ and $p=4$ circuits for the $N=3$ and $N=4$ states, respectively, achieve a fidelity of $F=0.95$ in noiseless simulation. The $N=3$ circuit contains $168$ $CZ$ gates with a 2-qubit gate depth $d=28$, while the $N=4$ circuit contains $264$ $CZ$ gates with a 2-qubit depth $d=44$. Both circuits were executed on a $4\times4$ patch of the 120-qubit IBM Nighthawk R1 processor (backend $\texttt{ibm\_miami}$).

The present optimization relies on classically accessible target states, limiting its scalability. However, the feasibility of the approach depends on the relative costs of classical target-state preparation, variational optimization, classical circuit simulation, and quantum circuit execution, which scale differently with system size. In the present work, we exploit classical resources to optimize shallow circuits for a regime in which the target states remain classically accessible. Beyond this regime, alternative strategies, including preparation protocols based on analytically constructed target states or variational approaches, may become more suitable, particularly when explicit knowledge of the target wave function is unavailable.

Beyond the limitations of classical optimization, scaling to larger lattices is also constrained by hardware fidelity and achievable circuit depth. In particular, as topologically ordered target states possess long-range entanglement, their preparation becomes increasingly challenging as the lattice grows, potentially requiring deeper circuits and more entangling operations. Moreover, finding a sufficiently large, well-connected patch of physical qubits with consistently high gate fidelity becomes more difficult, while an increased qubit number exposes the circuit to additional noise.

The principal limitation of the present experiment is hardware noise. For computational-basis measurements, particle-number post-selection retained $34.6\%$ of $80{,}000$ shots for $N=3$ and $33.3\%$ of $60{,}000$ shots for $N=4$. After post-selection, noise-scaling simulations estimate fractions of error-free realizations of $60(2)\%$ for $N=3$ and $46(5)\%$ for $N=4$. The remaining errors primarily attenuate the measured configuration probabilities and coherences, and cannot be removed by particle-number post-selection alone. Recovering unbiased observables will therefore become increasingly important when extending these simulations to regimes without classically available benchmarks. Improvements in gate fidelity, coherence, and sampling rate should make more demanding error-mitigation strategies practical.

The present platform provides a starting point for more direct probes of topological order and fractional excitations. In particular, the many-body Chern number could be accessed through randomized-measurement protocols that infer the invariant directly from statistical correlations of the prepared wave function~\cite{Cian21, Bringewatt26}, while topological entanglement entropy would provide a complementary nonlocal diagnostic of the FCI phase. The ability to prepare localized quasihole excitations further opens the possibility of manipulating their positions and extracting fractional braiding phases interferometrically. Extending these protocols to larger systems and particle numbers would ultimately allow digital quantum processors to address equilibrium and nonequilibrium properties of fractionalized states beyond the reach of exact classical simulation.

\acknowledgments{T.G. and K.D. acknowledge support from the Department of Science, Universities, and Innovation of the Basque Government through the BasQ strategy (project EMISGALA) and the IKUR Strategy (project QT8 RyGaQuS), and from the Agencia Estatal de Investigación (AEI) through Proyectos de Generación de Conocimiento PID2022-142308NA-I00 (EXQUSMI) and PID2025-175305NB-I00 (AQUASFONDAS).
K.D. acknowledges the support of the ``la Caixa'' Foundation.
The project that gave rise to these results
received the support of a fellowship from
the ``la Caixa'' Foundation (ID
100010434). The fellowship code is
LCF/BQ/DFI25/13000067.
B.A. acknowledges funding from the Maison du Quantique de Nouvelle-Aquitaine “HybQuant,” as part of the HQI initiative and France 2030, under French National Research Agency (ANR) Grant No. ANR-22-PNCQ-0002, and from Funda\c{c}\~ao de Amparo \`a Pesquisa do Estado de S\~ao Paulo (FAPESP), Grant No. 2018/25225-9.
The authors acknowledge the computational resources and technical support provided by the BasQ Strategy, under the collaboration agreement between Ikerbasque Foundation and Donostia International Physics Center (DIPC), on behalf of the Department of Science, Universities and Innovation of the Basque Government.
We acknowledge the use of generative AI tools for assistance with manuscript writing and code development. The authors take full responsibility for the content of this work.
}

\section{Methods}

\begin{figure*}[t]
\centering    \includegraphics[width=0.98\textwidth]{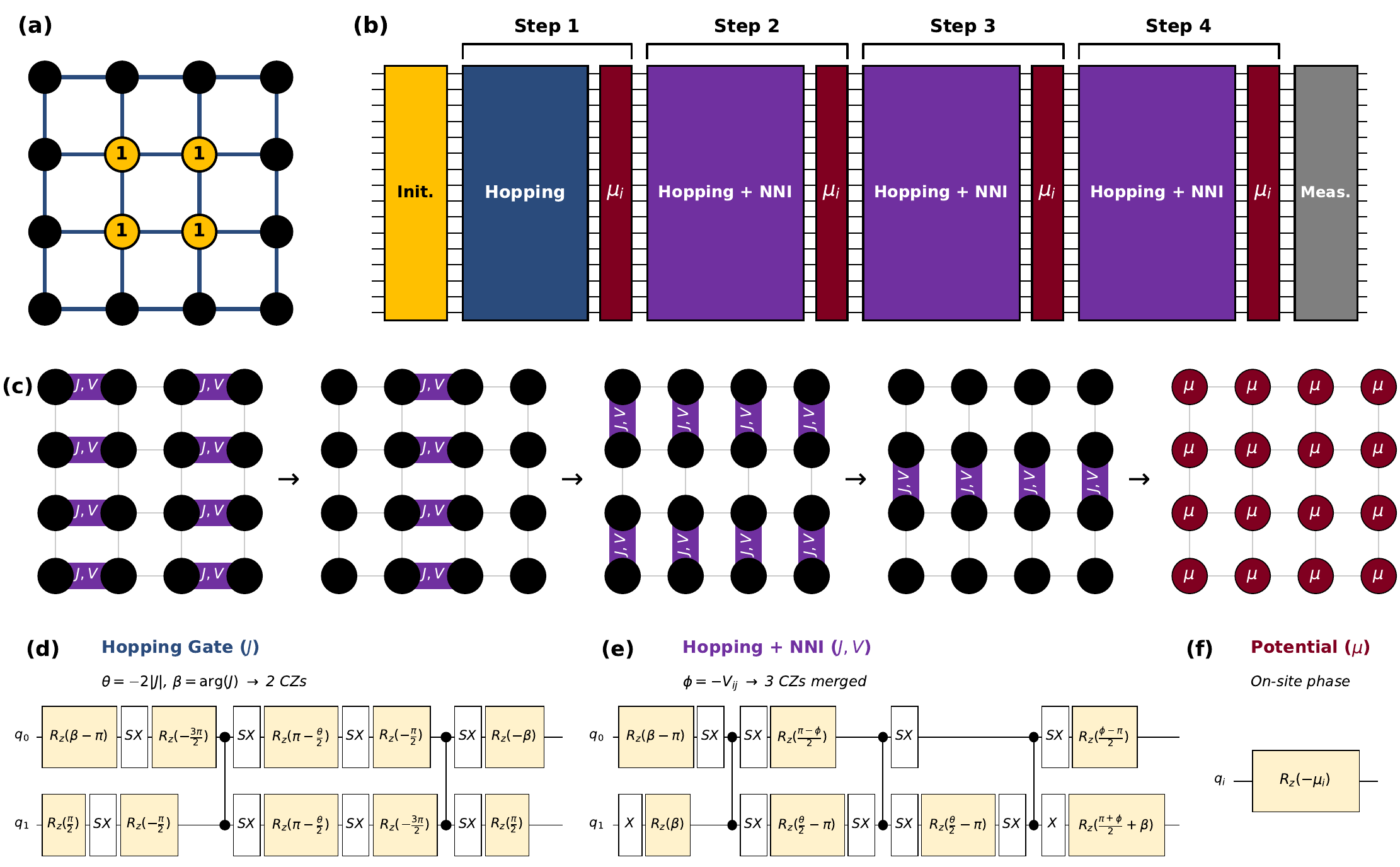}
   \caption{\textbf{State-preparation circuit and gate decomposition for the $N=4$ FCI state.}
(a) Initial product state, with yellow sites initialized in $\ket{1}$ and all remaining sites in $\ket{0}$. (b) Variational preparation protocol with $p=4$ steps, reaching a noiseless fidelity $F=0.95$; nearest-neighbor interactions are included in three of the four steps. (c) Decomposition of a single variational step into four layers of non-overlapping nearest-neighbor bonds that can be executed in parallel, followed by a layer of on-site potential rotations. (d)--(f) Gate implementations of a complex hopping amplitude $J$, a nearest-neighbor interaction $V$, and an on-site potential $\mu$, respectively. The hopping operation is implemented using Qiskit's \texttt{XXPlusYYGate}, requiring two $CZ$ gates, while inclusion of the nearest-neighbor interaction through a controlled-phase gate increases the cost to three $CZ$ gates per bond.}
    \label{fig:circuitAppendix}
\end{figure*}

\subsection{Kapit--Mueller parent Hamiltonian}
\label{app:km_ham}

The target FCI states are obtained by exact diagonalization of the interacting parent Hamiltonian introduced by Kapit and Mueller~\cite{Kapit10}. For bosons on a two-dimensional square lattice in a uniform magnetic field, its single-particle hopping Hamiltonian is
\begin{equation}
H_0=\sum_{j\neq k}J(z_j,z_k)\,a_j^\dagger a_k,
\end{equation}
where $z_j=x_j+iy_j$ denotes the position of lattice site $j$ in units of the lattice spacing. In the symmetric gauge, the hopping amplitude is
\begin{equation}
J(z_j,z_k)
=
W(z)\,
\exp\left[
\frac{\pi}{2}
\left(
z_j z^*-z_j^*z
\right)\phi
\right],
\end{equation}
with $z=z_k-z_j$ and $\phi=p/q$ the magnetic flux per plaquette in units of the flux quantum. Its spatial dependence is chosen as
\begin{equation}
W(z)
=
t\,(-1)^{x+y+xy}
\exp\left[
-\frac{\pi}{2}(1-\phi)|z|^2
\right],
\end{equation}
where $x=\operatorname{Re}(z)$ and $y=\operatorname{Im}(z)$. This fine-tuned long-range hopping produces an exactly flat lowest band whose single-particle states reproduce those of the lowest Landau level. We set $t=1$ throughout and use it as the unit of energy.

For a finite $L\times L$ lattice with magneto-periodic boundary conditions, the wave function satisfies
\begin{equation}
\psi(z+nL+imL)=\psi(z)e^{i\pi\phi L(ny-mx)}.
\end{equation}
The hopping amplitudes are obtained by summing over periodic images. Introducing lattice translations
\begin{equation}
R=nL+imL,
\qquad n,m\in\mathbb Z,
\end{equation}
the finite-system hopping amplitude is
\begin{equation}
J_L(z_j,z_k)
=
\sum_R
J(z_j,z_k+R)
\exp\left[
\frac{\pi}{2}
\left(
z_jR^*-z_j^*R
\right)\phi
\right].
\end{equation}

The many-body Hamiltonian used to obtain the target states consists of the interacting Kapit--Mueller parent Hamiltonian and local potentials $\mu_j$ used to pin the quasiholes:
\begin{equation}
H
=
\sum_{j\neq k}
J_L(z_j,z_k)\,
a_j^\dagger a_k
+
\frac{U}{2}
\sum_j n_j(n_j-1)
+
\sum_j\mu_j n_j,
\end{equation}
where $n_j=a_j^\dagger a_j$. We consider a $4\times4$ lattice at flux density $\phi=1/2$ and take the hard-core limit $U\rightarrow\infty$, restricting the local occupation to $n_j=0,1$. This allows each lattice site to be encoded in a qubit, with \(\ket{0}\) and \(\ket{1}\) representing the unoccupied and occupied states, respectively. The corresponding operator mapping is \(a_j\leftrightarrow\sigma_j^-\) and \(a_j^\dagger\leftrightarrow\sigma_j^+\), where \(\sigma_j^\pm\) are the qubit raising and lowering operators.

For the \(N=4\) case, we set \(\mu_j=0\) at all sites. At filling factor $\nu = 1/2$, the four bosons form the twofold-degenerate Laughlin-like ground-state manifold on the torus.

To obtain the two-quasihole target state, we reduce the particle number to $N=3$ and introduce strong local repulsive potentials at the desired quasihole positions. Specifically, we set $\mu_{q_1} = \mu_{q_2} = 100$ at the two pinning sites and $\mu_j = 0$ elsewhere. For both the $N=4$ ground state and the $N=3$ quasihole state, exact diagonalization yields two orthogonal states spanning the degenerate target manifold.

\subsection{Circuit construction and parameters}

The variational ansatz was designed to maximize gate parallelism while preserving total particle number. Each variational step $q$ is decomposed into four layers of non-overlapping nearest-neighbor bonds, as illustrated in Fig.~\ref{fig:circuitAppendix}(c), allowing all 2-qubit operations within a layer to be executed in parallel. These layers are followed by single-qubit $R_z(-\mu_i^q)$ rotations implementing the on-site potentials; $R_z$ rotations are native gates on the IBM Nighthawk R1 processor. The complex hopping amplitudes $J_{ij}^q$, nearest-neighbor interactions $V_{ij}^q$, and on-site potentials $\mu_i^q$ are optimized independently for each bond, site, and variational step.

Hopping between neighboring sites with complex amplitude $J$ for unit time is implemented using the parameterized \texttt{XXPlusYYGate} in Qiskit, with parameters
\begin{equation}
    \theta=-2|J|,
    \qquad
    \beta=\arg(J).
\end{equation}
The corresponding decomposition requires two $CZ$ gates together with single-qubit $SX$ and $R_z$ rotations [Fig.~\ref{fig:circuitAppendix}(d)]. A nearest-neighbor density interaction of strength $V_{ij}$ is implemented using a controlled-phase operation with phase
\begin{equation}
    \varphi=-V_{ij}.
\end{equation}
Combining hopping and interaction on the same bond yields three layers of $CZ$ gates, supplemented by $SX$, $R_z$, and $X$ gates [Fig.~\ref{fig:circuitAppendix}(e)]. The on-site chemical potential is implemented by the $R_z(-\mu_i)$ rotation shown in Fig.~\ref{fig:circuitAppendix}(f).

The high-level circuits were constructed in \textit{Qiskit} using \texttt{PhaseGate}, \texttt{XXPlusYYGate}, and \texttt{CPhaseGate}. They were compiled for the target processor with \texttt{generate\_preset\_pass\_manager} at \texttt{optimization\_level=3}, allowing the transpiler to synthesize and optimize the circuit in the native gate set consisting of the 2-qubit gate $CZ$ and the single-qubit gates $SX$, $X$, and $R_z$. A connected $4\times4$ patch of physical qubits was selected for the implementation. To reduce sensitivity to spatial variations in device performance, each circuit was executed using eight logical-to-physical mappings related by rotations and reflections of the virtual lattice. Experimental observables were averaged over these mappings, and their standard deviation was used to quantify mapping-dependent variations.

Experiments were executed using the \textit{Qiskit Runtime} \texttt{SamplerV2} primitive. Based on preliminary benchmarking of the available execution settings, we employed \texttt{XpXm} dynamical decoupling to suppress decoherence during idle periods, particularly for qubits at the corners of the lattice. Active qubit initialization was enabled together with a repetition delay of $4\,\mathrm{ms}$ to allow for thermal relaxation between successive circuit executions. Gate and measurement twirling were disabled. Data were acquired over multiple device-calibration cycles, with the number of shots for each observable specified in the corresponding measurement sections.

\subsection{Classical optimization}
\label{app:jax}

Classical optimization of the state-preparation circuits was performed using the JAX library~\cite{Bradbury18,Blondel22}. Because the variational circuit strictly conserves total particle number, all simulations were restricted to the fixed-$N$ Fock subspace, whose dimension is
\begin{equation}
    d_N=\binom{L^2}{N}.
\end{equation}
For the $4\times4$ lattice considered here, this gives $d_3=560$ for $N=3$ and $d_4=1820$ for $N=4$, substantially reducing the computational cost compared with simulations in the full $2^{16}$-dimensional qubit Hilbert space.

As illustrated in Fig.~\ref{fig:circuitAppendix}(c), each variational step consists of layers of mutually non-overlapping nearest-neighbor operations, together with an on-site potential layer. The evolution generated by each step Hamiltonian $H^q$ was evaluated numerically as $\exp(-iH^q)$ using a Taylor-series matrix-exponential algorithm with scaling and squaring, where the scaling factor and Taylor order were chosen adaptively according to the Frobenius norm $\|H^q\|_F$.

Each variational step has its own set of parameters: the complex hopping amplitudes \(J_{ij}^q\), nearest-neighbor interaction strengths \(V_{ij}^q\), and on-site potentials \(\mu_i^q\). All parameters were optimized jointly to maximize the manifold fidelity \(F\) defined in Eq.~\eqref{eq:fidelity}. The corresponding target state 
\psitgt is then obtained by projecting the optimized circuit output onto the target manifold (Eq.~\eqref{eq:psitgt}).

A step without nearest-neighbor interactions therefore contains $48$ real hopping parameters and $16$ on-site potentials, giving $64$ independent real parameters. Including the $24$ nearest-neighbor interaction strengths increases this number to $88$. The $N=3$ protocol uses $p=3$ steps, one of which contains nearest-neighbor interactions, resulting in
\begin{equation}
    K_{N=3}=2\times64+88=216
\end{equation}
variational parameters. The $N=4$ protocol uses $p=4$ steps, three of which contain nearest-neighbor interactions, giving
\begin{equation}
    K_{N=4}=64+3\times88=328.
\end{equation}

The large number of variational parameters makes finite-difference gradient estimation inefficient, since it requires a number of circuit evaluations that grows linearly with $K$. We instead used reverse-mode automatic differentiation in JAX, which evaluates the complete gradient at a computational cost comparable to a small number of forward evaluations and additionally enables GPU acceleration.

\begin{figure}
    \centering
    \includegraphics[width=0.45\textwidth]{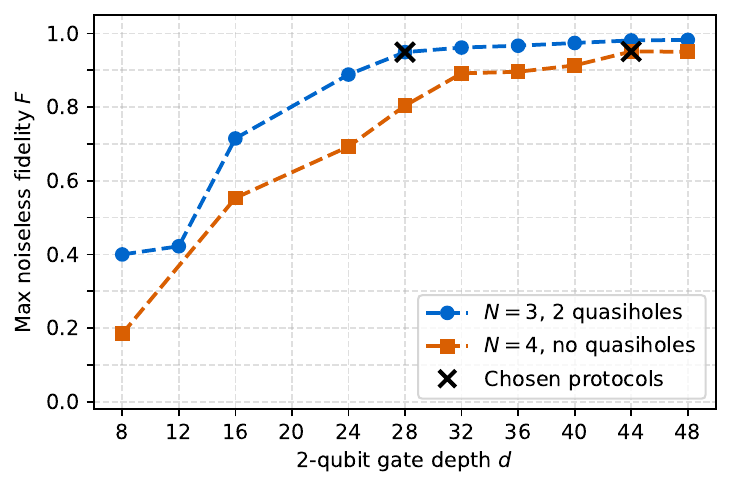}
    \caption{\textbf{Variational state-preparation performance.}
    Maximum noiseless fidelity $F$ as a function of 2-qubit gate depth for the $N=3$ state with two quasiholes (blue circles) and the $N=4$ state without quasiholes (orange squares). Each point shows the best result obtained from at least ten independent optimization runs for the corresponding circuit structure. Black crosses indicate the protocols selected for hardware execution, both reaching $F=0.95$.}
    \label{fig:fidDepth}
\end{figure}

For each circuit structure, the optimization was repeated from multiple initial parameter sets. Convergence was typically reached within $2000$ optimization iterations; the protocols ultimately selected for hardware execution converged after $1426$ iterations for $N=3$ and $1913$ iterations for $N=4$. Figure~\ref{fig:fidDepth} shows the maximum noiseless fidelity obtained as a function of the 2-qubit gate depth. Each point corresponds to the best result from at least ten independent optimization runs. Different ansatz structures can result in the same 2-qubit depth $d$ -- for example, three steps without nearest-neighbor interactions and two steps including nearest-neighbor interactions both have $d=24$. The $N=4$ target state requires a deeper circuit than the $N=3$ quasihole state to reach the same fidelity. The protocols chosen for hardware execution achieve $F=0.95$ with 2-qubit depths of $d=28$ for $N=3$ and $d=44$ for $N=4$.

\subsection{Measurement of bond currents}
\label{app:currents}
We reconstructed the bond currents associated with the Kapit--Mueller Hamiltonian from 2-qubit correlation measurements performed on the prepared $N=3$ quasihole state. The state-preparation circuit itself does not implement the Kapit--Mueller Hamiltonian dynamically; instead, the expectation values of its current operators are evaluated from the coherences of the prepared state. For a directed bond from site $j$ to site $i$, the current is

\begin{equation}
I_{j\rightarrow i}
=
2\,\operatorname{Im}
\left[
J_{ij}\langle a_i^\dagger a_j\rangle
\right],
\end{equation}
where $J_{ij}$ is the complex hopping amplitude of the Kapit--Mueller Hamiltonian.
For hard-core bosons, $a_i^\dagger=(X_i-iY_i)/2$ and $a_j=(X_j+iY_j)/2$, such that

\begin{equation}
a_i^\dagger a_j
=
\frac{1}{4}
\left(
X_iX_j+Y_iY_j
\right)
+
\frac{i}{4}
\left(
X_iY_j-Y_iX_j
\right).
\end{equation}
The real and imaginary parts of the hopping coherence are therefore reconstructed independently as
\begin{equation}
\operatorname{Re}\langle a_i^\dagger a_j\rangle
=
\frac{1}{4}
\left(
\langle X_iX_j\rangle
+
\langle Y_iY_j\rangle
\right),
\end{equation}
and
\begin{equation}
\operatorname{Im}\langle a_i^\dagger a_j\rangle
=
\frac{1}{4}
\left(
\langle X_iY_j\rangle
-
\langle Y_iX_j\rangle
\right).
\end{equation}

Writing $J_{ij}=J_{ij}^{\rm R}+iJ_{ij}^{\rm I}$, the current can equivalently be evaluated directly from the four measured Pauli correlators,

\begin{equation}
I_{j\rightarrow i}
=
\frac{1}{2}
\left[
J_{ij}^{\rm I}
\left(
\langle X_iX_j\rangle+\langle Y_iY_j\rangle
\right)
+
J_{ij}^{\rm R}
\left(
\langle X_iY_j\rangle-\langle Y_iX_j\rangle
\right)
\right].
\end{equation}

For each bond $(i,j)$, we collected $5{,}000$ shots for each of the four Pauli settings $XX$, $YY$, $XY$, and $YX$, measuring all remaining qubits in the computational ($Z$) basis. The measurements covered the 12 bonds forming the directed loop shown in Fig.~\ref{fig:Current}(a,b), for a total of $240{,}000$ shots before post-selection.

The simultaneous $Z$-basis readout of all sites outside the active bond allows particle-number post-selection to be applied to these measurements. For the $N=3$ state, we retained only outcomes satisfying

\begin{equation}
\sum_{k\neq i,j} n_k=N-1=2.
\end{equation}

Within the ideal fixed-$N$ sector, this condition selects configurations containing exactly one particle on the active bond, $n_i+n_j=1$. Configurations with $n_i+n_j=0$ or $2$ make no contribution to $\langle a_i^\dagger a_j\rangle$, since the hard-core hopping operator $a_i^\dagger a_j$ connects only the states $|01\rangle$ and $|10\rangle$ of the active bond. The post-selected Pauli correlators therefore yield the hopping coherence conditioned on the single-particle sector of the active bond and, correspondingly, a conditional bond current $I_{j\rightarrow i}^{\rm cond}$.

The unconditional physical current was recovered by weighting the conditional current with the probability of single occupation of the active bond,

\begin{equation}
I_{j\rightarrow i}
=
P(n_i+n_j=1)\,
I_{j\rightarrow i}^{\rm cond}.
\end{equation}
The bond-occupation probabilities $P(n_i+n_j=1)$ were determined independently for each bond from the computational-basis measurements of the particle-configuration distribution. Averaged over the $12$ measured bonds, this factor was
\begin{equation}
\langle P(n_i+n_j=1)\rangle =0.42,
\end{equation}
while the combined bond-wise post-selection retained on average $21.4\%$ of the measured shots.

To reduce sensitivity to spatial variations across the processor, the complete measurement protocol was repeated for eight physical-qubit mappings related by rotations and reflections of the virtual $4\times4$ lattice. Reported bond currents correspond to the mean over these mappings, and error bars denote their standard deviation. The total circulation around the quasiholes was obtained by summing the directed bond currents along the closed path $\mathcal C$,
\begin{equation}
I_{\rm total}
=
\sum_{(j\rightarrow i)\in\mathcal C}
I_{j\rightarrow i}.
\end{equation}

\subsection{Noise scaling analysis}
\label{app:noise_scaling}

\begin{table}[b]
\centering
\caption{\textbf{Hardware calibration parameters.} Calibration statistics across three hardware calibration cycles spanning the experiment for the 16 active physical qubits on \textit{IBM Nighthawk R1} (\texttt{ibm\_miami}). The reported 16-qubit layered $CZ$ error is $5.61\times10^{-3}$.}
\label{tab:calibration_summary_16q}
\begin{tabular}{lcccc}
\hline\hline
\textbf{Metric} & \textbf{Mean} & \textbf{Median} & \textbf{Min} & \textbf{Max} \\ \hline
$T_1$ ($\mu\mathrm{s}$) & $280.12$ & $286.78$ & $59.04$ & $485.75$ \\
$T_2$ ($\mu\mathrm{s}$) & $203.24$ & $202.89$ & $52.54$ & $389.19$ \\
$SX$ error ($\times10^{-4}$) & $3.40$ & $2.42$ & $0.79$ & $21.32$ \\
Isolated $CZ$ error ($\times10^{-3}$) & $5.89$ & $3.79$ & $2.23$ & $14.54$ \\
Readout error $p(0|1)$ (\%) & $1.15$ & $1.17$ & $0.61$ & $1.76$ \\
Readout error $p(1|0)$ (\%) & $0.76$ & $0.76$ & $0.12$ & $1.39$ \\
\hline\hline
\end{tabular}
\end{table}

\begin{figure}[t]
    \centering
    \includegraphics[width=\linewidth]{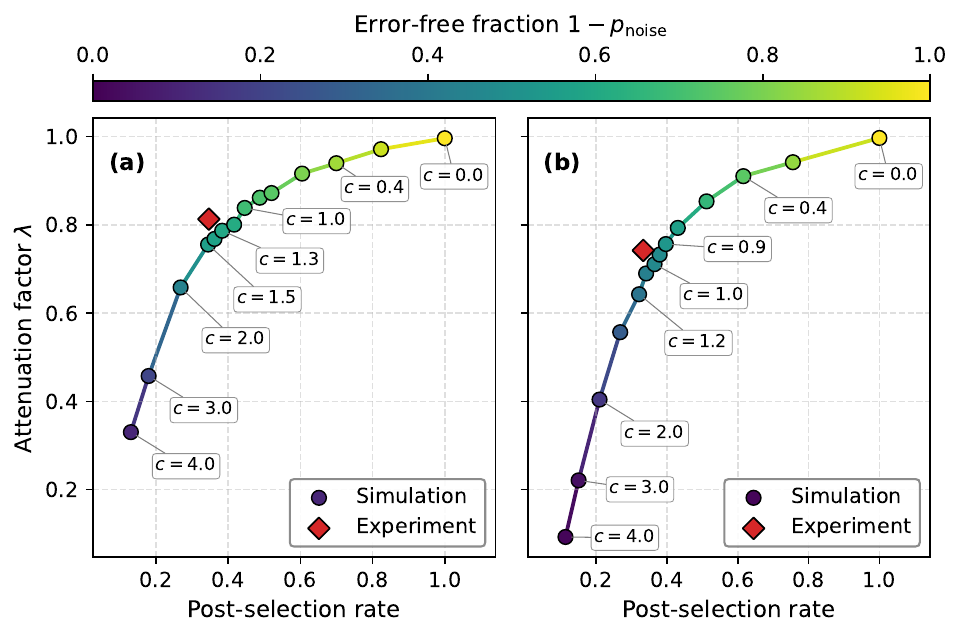}
    \caption{\textbf{Noise-scaling analysis and estimation of the error-free fraction.}
    Parametric trajectories of the attenuation factor $\lambda$ and particle-number post-selection fraction obtained from noisy \texttt{Qiskit Aer} simulations as the noise scale $c$ is varied from $0$ to $4$. Each simulated point is obtained from $4096$ trajectories with $20$ shots per trajectory, and the color scale indicates the corresponding error-free fraction $1-p_{\rm noise}$. Experimental results (red diamonds) lie closest to the simulated trajectories near $c=1.3$ for $N=3$ [panel (a)] and $c=1.0$ for $N=4$ [panel (b)], corresponding to estimated error-free fractions of approximately $0.60$ and $0.46$, respectively.}
    \label{fig:noise_scaling}
\end{figure}

Table~\ref{tab:calibration_summary_16q} summarizes the calibration data for the 16 active physical qubits over the three calibration cycles spanned by the experiment. Isolated 2-qubit calibration errors characterize individual couplers and have a median $CZ$ error of $3.79\times10^{-3}$. Because the state-preparation circuits execute several 2-qubit gates simultaneously, we instead use the IBM-reported 16-qubit layered $CZ$ error of $5.61\times10^{-3}$ to parameterize the 2-qubit noise. This metric incorporates the degradation associated with parallel gate execution and therefore more closely reflects the operating conditions of our circuits.

We simulated the experimental circuits in \texttt{Qiskit Aer} using a custom noise model. The baseline model, corresponding to noise scale $c=1$, was constructed from representative calibration parameters of the active qubits listed in Table~\ref{tab:calibration_summary_16q}. Single-qubit physical gates ($SX$ and $X$) include thermal relaxation during a gate duration $t_{\rm 1q}=35\,\mathrm{ns}$, using $T_1=287\,\mu\mathrm{s}$ and $T_2=203\,\mu\mathrm{s}$, together with a single-qubit depolarizing error $\epsilon_{\rm 1q}=2.42\times10^{-4}$. On the Nighthawk processor, $R_z$ rotations are implemented virtually as phase-frame updates, requiring no physical pulses. Accordingly, we assign them no gate duration or gate error in our noise model. 2-qubit operations are modeled using a depolarizing channel with error rate $\epsilon_{\rm 2q}=5.61\times10^{-3}$, corresponding to the measured layered $CZ$ error. Asymmetric readout noise is applied after sampling using the state-dependent probabilities $p(0|1)=1.17\%$ and $p(1|0)=0.76\%$.

To explore different overall noise strengths, the gate depolarizing probabilities and readout-error probabilities are scaled linearly according to
\begin{equation}
    \epsilon(c)=c\,\epsilon_0,
\end{equation}
where $\epsilon_0$ denotes the corresponding baseline value at $c=1$. The coherence times $T_1$ and $T_2$ are kept fixed at their hardware values.

To relate the attenuation observed in the configuration probabilities to the fraction of error-free realizations, we first consider a simple global-depolarizing model,
\begin{equation}
    P_{\rm exp}
    =
    (1-p_{\rm noise})P_{\rm sim}
    +
    p_{\rm noise}P_{\rm rand},
    \label{eq:noise_mixture}
\end{equation}
where $P_{\rm sim}$ is the noiseless circuit distribution, $P_{\rm rand}$ is the maximally mixed distribution within the fixed-$N$ particle sector, and $p_{\rm noise}$ is the residual error probability after particle-number post-selection. At the level of individual fixed-$N$ bitstrings, $P_{\rm rand}$ is uniform. After grouping configurations into symmetry classes, its probability varies according to the degeneracy of each class.

Subtracting $P_{\rm rand}$ from Eq.~\eqref{eq:noise_mixture} gives
\begin{equation}
    P_{\rm exp}-P_{\rm rand}
    =
    (1-p_{\rm noise})
    \left(P_{\rm sim}-P_{\rm rand}\right).
\end{equation}
Within this idealized model, the slope $\lambda$ of a linear fit of $P_{\rm exp}-P_{\rm rand}$ against $P_{\rm sim}-P_{\rm rand}$ therefore directly equals the error-free fraction,
\begin{equation}
    \lambda=1-p_{\rm noise}.
\end{equation}
The attenuation factors reported in the main text are obtained from the symmetry-grouped configuration probabilities.

Realistic noise does not, however, produce a purely maximally mixed background after post-selection. For example, pairs of bit-flip errors can conserve the total measured particle number while effectively displacing a particle. Such events can preserve part of the spatial structure of the remaining particles and therefore remain positively correlated with the noiseless distribution. Consequently, the residual-error distribution differs from $P_{\rm rand}$, and the measured attenuation factor $\lambda$ generally overestimates the true fraction of error-free realizations.

We quantify this effect by simulating the experiment over noise scales $c\in[0,4]$. For each value of $c$, we use $4096$ simulated trajectories with $20$ shots per trajectory and determine the particle-number post-selection fraction, the attenuation factor $\lambda$, and the fraction $1-p_{\rm noise}$ of trajectories containing no sampled error event. Figure~\ref{fig:noise_scaling} shows the resulting trajectories in the $\lambda$--post-selection plane, with the color scale indicating $1-p_{\rm noise}$. The experimental values lie closest to the simulated trajectory near $c=1.3$ for $N=3$ and $c=1.0$ for $N=4$. Matching both the measured attenuation and post-selection fraction gives estimated error-free fractions
\begin{equation}
    1-p_{\rm noise}=0.60(2)
    \quad (N=3),
\end{equation}
and
\begin{equation}
    1-p_{\rm noise}=0.46(5)
    \quad (N=4).
\end{equation}
The lower error-free fraction for $N=4$ is consistent with its greater circuit depth and correspondingly larger accumulated error probability.

These estimates depend on the assumed noise model. In particular, we use representative rather than site-resolved calibration parameters, and the depolarizing model does not capture coherent, state-dependent crosstalk or spectator-qubit dynamics during dense parallel gate execution. Moreover, scaling gate and readout errors through a single parameter $c$ does not allow their contributions to be fitted independently. Nevertheless, the proximity of the experimental data to the simulated trajectories around the hardware baseline $c\simeq1$ indicates that the model captures the dominant attenuation and post-selection behavior. The quoted uncertainties are obtained from the variation of $1-p_{\rm noise}$ between neighboring values of $c$ around the experimental match and therefore quantify only the uncertainty within this noise model; systematic modeling errors may be larger.

\end{document}